\documentclass[manuscript]{acmart}

\setcopyright{none}
\copyrightyear{2026}
\acmYear{2026}
\acmDOI{XXXXXXX.XXXXXXX}

\usepackage{todonotes}

\begin{document}

\title{Mapping justice as a plural and contested concept in human--computer interaction}

\author{Henrik Rydenfelt}
\authornote{Both authors contributed equally to this research.}
\email{henrik.rydenfelt@gmail.com}
\orcid{0000-0003-3739-4712}
\affiliation{%
   \department{Faculty of Theology}
  \institution{University of Helsinki}
  \city{Helsinki}
  \country{Finland}
}

\author{Matti Nelimarkka}
\authornotemark[1]
\orcid{0000-0001-9867-692X}
\affiliation{%
    \department{Social Computing Group}
    \department{Department of Computer Science}
    \institution{University of Helsinki}
    \city{Helsinki}
    \country{Finland}}
\affiliation{%
    \department{Department of Computer Science}
    \institution{Aalto University}
    \city{Espoo}
    \country{Finland}}
\email{matti.nelimarkka@helsinki.fi}

\begin{abstract}

Human–computer interaction increasingly studies and promotes justice.
Often, this work does not treat justice through a pluralistic lens
Justice, however, is not one thing: political philosophy offers several well-established, mutually incompatible conceptions of it.
To navigate these meanings, we draw on a map of justice, a philosophical framework identifying six distinct conceptions of justice situated throughout (Western) philosophical thought.
We apply this framework to four cases that illustrate different approaches to social justice: the COMPAS system, food democracy, gentrification, and robotics.
Our analysis shows that each case already reflects a position on the map, but more importantly, one might adapt the case to other conceptions as well, leading to very different systems and analyses.
We discuss how HCI might engage more carefully with such contested concepts, and warn that without pluralistic understanding, academic discourse risks a spiral of silence.
\end{abstract}

\maketitle

\section{Introduction}

Over the past two years, \textit{The Economist} has claimed to trace a shift in public debate about justice.
In 2024 it pointed to a decline in the use of words such as diversity, equity, inclusion and white privilege -- markers of a societal perspective focused on social justice -- in outlets such as the \textit{Washington Post}, and diagnosed this as the passing of what it called ``peak woke''.\footnote{~The Economist, Sep 19th 2024, \url{https://www.economist.com/leaders/2024/09/19/after-peak-woke-what-next} and \url{https://www.economist.com/briefing/2024/09/19/america-is-becoming-less-woke}.}
In 2026, ,the newspaper went on to claim that the terms of the debate have shifted with a title: ``Woke 1 was about identity. Woke 2 is about class.''\footnote{~The Economist, Sep 3rd 2026, \url{https://www.economist.com/united-states/2026/09/03/woke-1-was-about-identity-woke-2-is-about-class}.}
A shift is not only in the newspapers, but also in the political climate.
For example, the Trump administration has been rapidly conducting actions which have stepped on civic liberalities and the role of state -- in addition to actions they have taken to dismantle the democratic system and causing democratic backsliding -- and in Europe conservative or conservative-leaning parties and governments have gained new ground -- for example, the far-right Alternative für Deutschland gained 43\% support in the 2026 Saxony-Anhalt (German state) election.
Such changes bring justice again under open dispute.

Human--computer interaction researchers have engaged justice on many fronts: designing technology to further it,
and using it as a lens on socio-technical arrangements \citep{Boone2024,Bellini2022,Ohlund2025}.
At the same time, different people can justifiably claim to work toward a just society while meaning considerably different things by it -- and interpreting, accordingly, what such work entails.
A design or a research programme built on one conception will be met by people who hold the others, whatever the political climate of the moment.
This highlights the need to understand this complexity of competing values and interests-

To navigate this complexity, we adapt a framework from ethics and political philosophy \citep{hayry2018doctrines,hayry2021just,rydenfelt2024climate,rydenfelt2021justice} that systematically distinguishes between six competing conceptions of justice, clearly illuminating the structure of disagreements among thems.
This framework invites human--computer interaction scholars to consider \textit{what kind of society they are after} and \textit{how their ideal society differs from those based on alternative conceptualisations of justice}.
It invites our research community to be more explicit about the commitments and their underpinnings involved in critical and value-driven work, and to acknowledge the existence of different value perspectives.

Our discussion builds on conceptual work and provides illustrative examples of how different justices manifest in technology.
We first review existing literature on justice in HCI, and then introduce our framework.
Following this, we illustrate how this framework impacts both the design and evaluation of novel technologies.
We conclude with a discussion of the implications for the field. 
However, given the normative nature of this subject, we first provide a brief positionality statement to reflect on how our own cultural and disciplinary backgrounds have shaped this inquiry.

\subsection*{Author positionalities}

Given the nature of our work, we believe that it is important for us to make explicit our background and motivations to explore justice as a concept in the context of human-computer interaction.

The first-named author's interest in the 'pluralism of justice' is rooted in experience navigating the gap between US-centric academic discourse and the realities of a Western European social democracy, emerging from his background in political science and human--computer interaction.
While supporting the core tenets of social justice-oriented HCI, the author is cautious of the field’s tendency to export US-based cultural assumptions to contexts with different social and political histories and institutional trust structures.
Drawing on a background in political science, the author views democracy as a pluralistic negotiation between competing visions of the 'good life' or `just society.'
Consequently, they are concerned that when socio-technical arrangements cluster around a narrow set of values, they may inadvertently simplify the richness of social considerations, potentially leading to a culture of silence regarding alternative societal arrangements.

The second-named author's background in philosophy and the social sciences has raised a persistent concern with the use of `justice' in research and design without an account of what the word is taken to mean.
This is not a demand that every paper carry a theory of justice, but a concern about what the omission produces: when the meaning goes unstated, some conception fills the gap by default, and the result risks being parochial -- an implicit conception of justice taken as given.
The author's commitment in this work is accordingly not to any one conception on the map we present, including those currently prominent in the field, but to the questions that the map enables posing.

It is vital to note that while we are critical of how the term `justice' is used, we do not object to research that is driven by societal concerns or is activist in nature.
Rather, this essay navigates the complexities surrounding contested concepts such as justice with the aid of the map of justice we outline.
Our conceptual work and analysis aim to be non-normative: we do not intend to promote any particular conception of justice, but rather to show how different conceptions of justice, rooted in Western thought, lead to different questions and design directions. 

\section{Related work}

There are many ways to understand justice.
We first examine how social justice -- a coherent body of literature -- has been examined in human-computer interaction research.
Following that adapt the Map of Justice from political philosophy to help us navigate the meanings of justice.

\subsection{Justice in human--computer interaction research}

Papers using the term `social justice' have increased at CHI and DIS during the 2020s \citep{Boone2024,Bellini2022,Ohlund2025}.\footnote{As a related concept, (algorithmic) fairness has been extensively discussed in human--computer interaction  as well \citep{Mulligan2019,Selbst2019,Kasy2021,Corbett-Davies2018}.
Maybe the most critical observation in this discussion is the acknowledgement that different definitions of justice are mutually exclusive \citet{Lipton2017}, that is, one cannot achieve both disparate impact and disparate treatment fairness at the same time.}
The meaning of (social) justice has been discussed in workshops \citep{Bellini2022} and through systematic reviews of the literature \citep{Boone2024,Ohlund2025}, which identify a range of emphases and areas of focus, from social and racial justice to reproductive, intergenerational, environmental, and economic justice.
Research has also explored justice within academia, such as citational justice, as well as justice in wider society, such as restorative justice.
This body of work is underpinned by attention to systems of power, oppression, and privilege, and to the diverse forms of harm through which injustice can manifest, drawing from works of \citet{Collins2000}, \citet{Crenshaw1991} and \citet{Young1990} among others.
These harms range from physical and economic harms to emotional, cultural, and environmental harms, as well as harms associated with reduced opportunities for self-actualisation and autonomy \citep{Boone2024}.



In a systematic analysis of the literature, \citet{Ohlund2025} identifies four different streams of social justice oriented works:
design-oriented works on social justice as a design opportunity,
examinations of existing socio-technical systems with a social justice lens,
works for technical and professional communication (TPC) emerging around \citet{Walton2019-ru} work in the TPC community,
and
social justice in computer science and engineering, with a focus on pedagogical approaches to acknowledge fields' growing responsibility toward complex social challenges.
We focus on the distinction between design-oriented and action-driven research and research where social justice is used as an analytical lens; the distinction has also been identified in other literature reviews \citep{Boone2024}.

The design-focused stream leans heavily on a \textit{social justice-oriented interaction design} --approach for human--computer interaction scholars seeking to address large-scale social issues through technology design and development.
\citet{Dombrowski2016} call out three commitments essential for social justice-oriented design: conflict, reflexivity and personal politics and ethics.
For \citet{Dombrowski2016}, social justice entails acknowledging the diversity of people's experiences, which may in turn give rise to conflicts and tensions during the design process.
It also requires recognising designers' positionality and resisting the view of the designer as an objective or neutral actor. 
More fundamentally, social justice is understood as inherently political and ethical:
it entails a particular understanding of society and its problems, as well as commitments concerning how those problems ought to be addressed.
Building on these three commitments, \citet{Dombrowski2016} identify six design strategies for conducting social justice-oriented design, drawing on \citep{lotter2011poverty}.
For example, they advocate designing for transformation, which acknowledges that social relations are continually evolving and can shift attention from immediate innovations towards the structural conditions underlying inequality.
Similarly, they argue that social justice-oriented design should attend to equitable distribution, which they later refer to as `just distribution'.
Such distribution, they argue, should not only be equitable but also sustainable, and should be achieved through equitable decision-making processes.


The second stream on analytical focus uses social justice to ``as a lens to explicitly frame complex social challenges such as but not limited to, poverty, trans rights, class, fat positivity, racism, older adults rights, sex-worker rights and gentrification'' \citep{Ohlund2025}.
Unlike a design objective, justice is used to explore and understand social challenges, to power societal analysis and critique with a conceptual tool, and then situate technology and its impacts into the analysis and critique.
These works can, for example, discuss how technology serves as a new mechanism for oppression in light of conceptions of social justice.
As another example, social justice has been raised as a concept central to the examination of participation and issues concerning the inclusion of marginalised communities.

These examples show that the social justice within human--computer interaction is a broad umbrella under which researchers have been addressed and often sought to resolve societal issues. 
However, this research focus on social justice, with power, oppression, and privileged as key perspectives -- according to \citet{Ohlund2025} justice is a plural concept.
The particular direction from which the this body of literature has approached the concept becomes visible when considering the broader range of ways in which justice has been understood.

\subsection{The Map of Justice}

For our analysis, justice concerns what people are owed and what they may legitimately claim from one another and from social institutions. 
Justice is commonly associated with \emph{impartiality}:
the idea that similar cases should be treated in the same way can be traced back to \citet{aristotle}.
However, impartiality does not imply that everyone should receive the same treatment or the same share.
Different treatment may be justified where there is a relevant difference between the cases.

\begin{figure}[b]
    \centering
    \includegraphics[width=.75\linewidth]{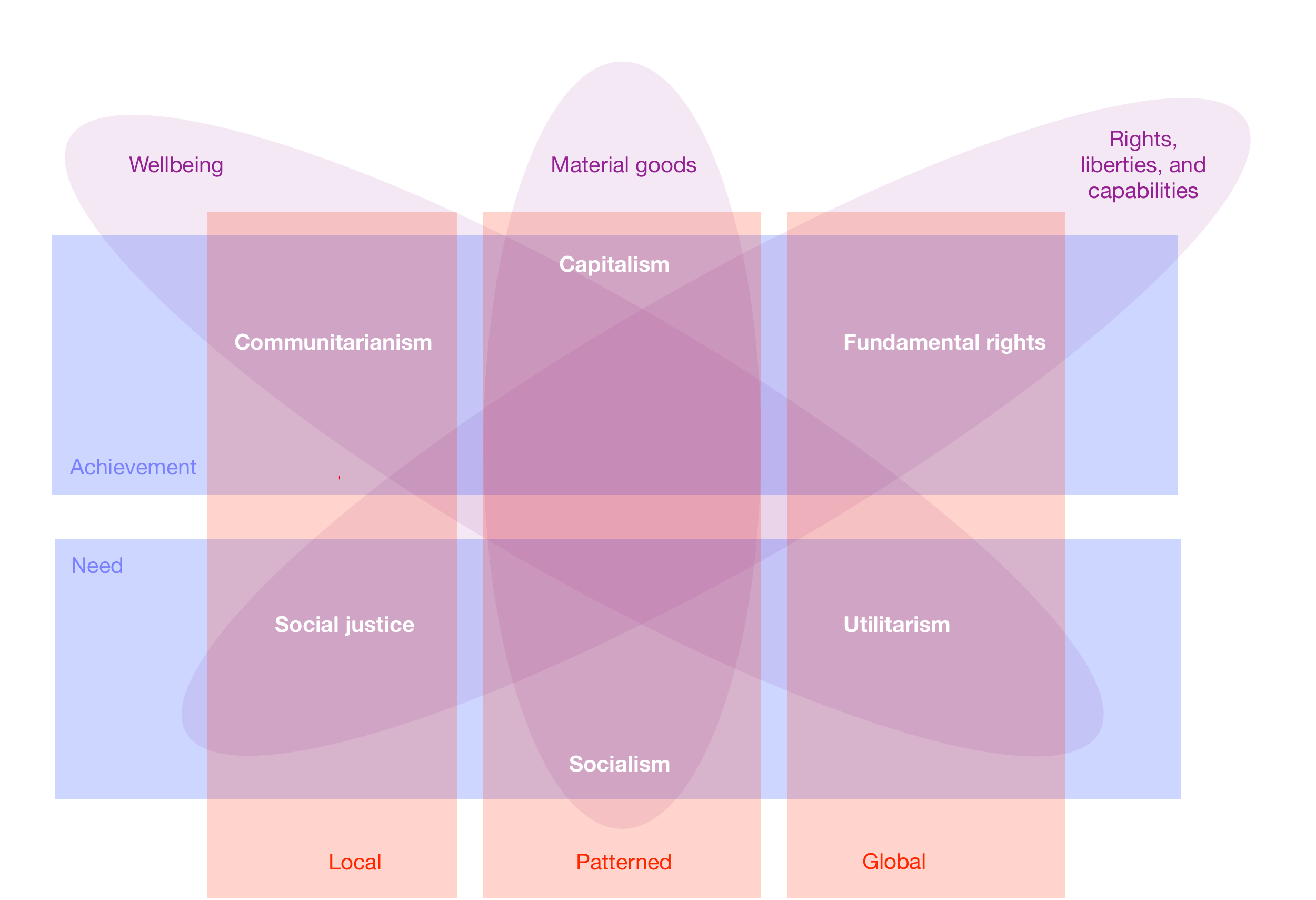}
    \caption{The Map of Justice, distinguishing the six different conceptions of justice (in white), and three dimensions they are focused on: the basis of distribution (blue, achievement and need), recipients of distribution (red, local, patterned and global) and the goods distributed (dark blue, wellbeing, material goods, and rights, liberties and capabilities).
    The map illustrates how conceptions of justice may share some principles with other conceptions while disagreeing with them on another axis.} 
    \Description{
    The conceptions of justice can be seen in relationship to each other based on three dimensions illustrated in the drawing.
    The distribution on Achievement is shared with Communitarianism, capitalism and Fundamental rights,
    while Distribution on Need is shared with Social justice, Socialism and Utilitarianism.
    Communitarianism  and social justice focus on local recipients of distribution,
    Capitalism and socialism on patterned recipients of distribution,
    and Fundamental rights and Utilitarianism on global distribution.
    Lastly, conceptions differ in what they considered to be distributed:
    capitalism and socialism focus on material goods,
    social justice and fundamental rights focus on rights, liberties and capabilities,
    and Communitarianism and Utiliarism focus on wellbeing.
    }
    \label{fig:map_of_justice}
\end{figure}

\emph{Distributive} justice concerns how benefits, burdens, opportunities, and other goods should be allocated among people.
To make the wide diversity of different accounts of justice tractable, we draw on a conceptual map of justice developed by the philosopher \citet{hayry2018doctrines,hayry2021just} and substantially extended by Rydenfelt \citep{rydenfelt2024climate,rydenfelt2021justice,extreme_justice}.
It condenses major strands of Western thought on distributive justice into six main conceptions (see Figure~\ref{fig:map_of_justice}).\footnote{Naturally, we acknowledge that there are various approaches to justice, also non-Western \citet{sen2009idea,ubuntu} and more-than-human accounts \citep{Nussbaum} on justice. Therefore, the Map of Justice is a selection of some perspectives on justice, and further analysis is welcomed to account for even further accounts.}
Before applications to technology domain, the map has proven useful across a range of settings: it has been applied to public debates during the COVID-19 pandemic \citep{hayry2021just,rydenfelt2021justice}, to questions of ecological sustainability \citep{hayry2025oikeudenmukainen}, to
environmental and climate policy debates where different justice claims pull in different directions \citep{rydenfelt2024climate}, and to discussions of the Nordic welfare state's development and future \citep{hyvinvointivalitio}.
We first describe the six conceptions, and following that, show the three dimension of the concept of justice which these conceptions highlight.

\paragraph{Capitalism}
On this map, capitalism names the view that a just distribution is whatever emerges from a free market -- one that allocates goods according to individual choice and achievement.
This view is tied to a strong defense of private property, often traced to classical liberalism, particularly \citet{locke}, who argued that labor grounds a natural right to property.
Contemporary libertarians such as \citet{nozick1974anarchy} fall under this heading: for Nozick, whatever distribution results from free, voluntary exchange between consenting parties is thereby just.\footnote{Nozick himself \citep[pp.~160--164]{nozick1974anarchy} rejected the term ``distributive justice,'' seeing it as smuggling in the idea that goods can be redistributed independently of how they were produced -- production being, for him, the real basis of entitlement.}
The basis of a just distribution is individual achievement, and the goods at stake are typically material goods that can be acquired and traded through the market.

\paragraph{Socialism}
Socialism, by contrast, holds that a just society distributes goods according to need.
Its clearest statement is Marx's description of a communist society: goods flow ``from each according to his ability, to each according to his need'' \citep{marx}.
Marx did not frame this as a theory of justice, but the phrase has nonetheless become a touchstone for later socialist accounts of justice \citep{gilabert2015socialist}.
The goods at stake are typically material goods, basic necessities and resources whose production and distribution are to be organized collectively, often through the state.

\paragraph{Fundamental rights}
Contemporary fundamental rights accounts draw on the theory of justice as fairness by \citet{rawls1971theory}.
Discussion of Rawls tends to gravitate toward his second principle -- that inequality is just only when it benefits society's worst-off -- but the more foundational claim is his first principle: that everyone has an equal claim to equal basic rights and liberties, including of political participation, compatible with the same scheme for all. Just institutions exist to secure these basic rights and liberties for all.

\paragraph{Utilitarianism}
Utilitarianism holds that the right action is whichever produces the most good overall. Early utilitarians like \citet{bentham} and \citet{mill} equated ``good'' with happiness or utility; most contemporary utilitarians instead speak more broadly of well-being, assessed impersonally, from no particular person's point of view. As a theory of justice, utilitarianism says a distribution is just when it maximises total well-being.

\paragraph{Communitarianism}
Communitarianism has long historical roots.
Its contemporary versions developed largely as a reaction against Rawlsian liberalism.
Rawls's critics such as \citet{sandel} and \citet{walzer} argued that liberal theory neglects how deeply people's identities and values are shaped by the communities they belong to.
On this view, shared community traditions -- not abstract individuals -- are the proper starting point for ethical and political reasoning; the basis of a just distribution, and what counts as a relevant good, should be defined relative to a particular community's own values and practices.

\paragraph{Social justice}
As the term is used on this map, social justice refers to a contemporary cluster of views concerned with inequality -- especially in connection of marginalised or minority groups -- and should be distinguished from the broader, everyday sense of ``social justice'' that could describe any of the six conceptions here. In this narrower sense, social justice refers to a view of a just society as one where groups, especially marginalised ones, can voice their different needs and have those needs met. These accounts tend to be sceptical of one-size-fits-all, universalist views.
Key approaches falling under this label include
ethics of care \citep{gilligan1982different},
structural injustice \citep{young1990justice}, and
justice that combines redistribution with recognising individuals and groups as full participants in social life \citep{fraser2000rethinking}.
What unites these otherwise divergent accounts is the conviction that justice depends on securing the conditions under which people can articulate their own needs.

The map's analytical power is that it makes six rival positions comparable along three dimensions by asking the same three questions of each: \textit{on what basis} goods are owed, \textit{to whom}, and \textit{what} is owed (Figure \ref{fig:map_of_justice}):

\begin{description}
    \item[Basis:] 
    Above the horizontal axis, \emph{achievement} is the basis of distribution: communitarianism measures it against a community’s values, capitalism through market outcomes, and fundamental rights through being human itself.
    Below it, \emph{need} is the basis of distribution : social justice lets groups define their own needs, socialism views them as those of the individual, typically assessed collectively, and utilitarianism attempts to measure them impersonally. 
    \item[Recipients:]
    Communitarianism and social justice are \emph{local} in their distribution, with goods owed to members of a particular group or community;
    capitalism and socialism are \emph{patterned}, distributing goods to individuals, typically within a society;
    and fundamental rights and utilitarianism are \emph{global}, aiming at a just distribution for everyone.
    \item[Goods:]
    Capitalism and socialism focus on \emph{material goods} in their distribution; fundamental rights and social justice on \emph{rights, liberties, and capabilities}, including recognition and participation in social and political life; and utilitarianism and communitarianism on utility or \emph{well-being}.
\end{description}

\section{Illustrative examples}

To make the relevance of the six conceptions of justice more concrete, we next consider them in connection of examples drawn from justice oriented debates in human--computer interaction.
Our first example draws on the debate on the COMPAS system, a much discussed system studied in the critical computing literature \citep[among many others][]{Mulligan2019,corbett2017algorithmic,Paakkonen2020,christin2017algorithms,Alkhatib2019}.
The following two cases are chosen from \citep{Ohlund2025} as exemplars of design-oriented work (food democracy) and use of social justice as an analytical lens (gentrification).
Our final example on robots for social justice presents a case in which the concept of social justice itself is not explicit.
\citep{Ohlund2025} highlight that the case ``never bring[s] in any specific theoretical framework for social justice or providing any larger defining frame'', indicating that the while social justice is used as a term, it remains open-ended.
The last case allows us to examine social justice as almost-floating signifiers, losing its specific meaning -- a problematic way of using terms in human--computer interaction research \citep{Marshall2017}.
With each of examples, we illustrate considerations and aims from the six different conceptions of justice, showing that the the whole system would be different across these cases.

\subsection{COMPAS system}

The COMPAS system is used for risk-assessment of offenders recidivism likelihood in the criminal justice to aid judges to consider if an inmate is sent to jail or prison, i.e., if they are in the risk of continuing criminal behaviour \citep{christin2017algorithms}.
\citet{angwin2016machine} made a compelling case was made racially biased: it led to more high likelihoods for some races.
The various scholarly reinterpretations of this case have focused on fairness and domination \citep{Mulligan2019,corbett2017algorithmic,Burrell2024},
examining the roles of humans and algorithmic systems in this operation \citep{Paakkonen2020,christin2017algorithms},
or examined what types of bureaucracy emerge when algorithmic systems are used \citep{Alkhatib2019,Paakkonen2020}.

Rather than revisiting COMPAS's shortcomings, we ask how a system supporting such decisions might be designed under each of the six conceptions of justice.
The system recommends decisions on a person's physical liberty -- but it does so as part of a society.
The map's three questions thus become design questions: \textit{what} goods are at stake in the system's operation, on what \textit{basis} its recommendations should rest, and \textit{who} should be considered in its design.
Each conception answers these differently.

From a \textbf{capitalist} perspective, physical liberty becomes a good tradable on a market.
A just distribution rests on individual achievement, measured through the market: what a person has earned and what they are willing to pay.
Rather than predicting recidivism, the system would price liberty and process the exchange.
This is not far-fetched: the US bail system already forms a market in pre-trial liberty, where a defendant may pay the court a bond to be released before trial.

From a \textbf{socialist} perspective, the basis of a just distribution shifts from achievement to need.
Detention suspends the needs of daily life -- to work and earn a living, to maintain a home, to keep up the relationships one lives by.
The needs considered are often not limited to the detained person's own: a single parent's detention leaves a child without care.
Rather than predicting recidivism, the system would assess these needs and how they are to be met: sometimes by release, sometimes in other ways alongside detention.

From a \textbf{fundamental rights} perspective, physical liberty is a right held equally by all.
Liberty may be taken away only through an equal and contestable process: the person must be able to know, understand, and challenge the grounds on which their detention rests.
Everyone subject to such a system also holds an equal right to participate in deciding the rules it encodes.
A statistical prediction fails these demands twice over.
First, an opaque score cannot be examined or contested.
Second, a prediction of future behaviour cannot be challenged at all:
one can contest what one is accused of having done, not what one is calculated likely to do.
Under this conception, there may be no system of this kind to design at all.
If one were designed, it would support legal institutions in due process: stating the grounds of detention, keeping its records, and hearing its challenges.

From a \textbf{utilitarian} perspective, a just society maximises overall wellbeing, assessed impersonally.
A system designed on this basis would weigh the harms of detention to the individual against the harms incurred by release.
Under this conception, recidivism prediction finally becomes a proper part of the system's core.
Of the six, this is the conception COMPAS comes closest to embodying: its risk scores estimate one side of the balance, the harms of release.
Indeed, some of the criticism of COMPAS focused on its errors in performing this function, as opposed to questioning the undergirding goal of maximising overall wellbeing.

From a \textbf{communitarian} perspective, justice is grounded in the values and practices of a particular community.
What detention is for and when it is warranted are not questions with a single answer: communities differ in how they respond to wrongdoing and in what they expect of their members.
This conception, then, does not say what the system would decide -- it says where the system's standards must come from: the community whose members it concerns.
The community may be a whole society, whose criminal justice embodies a penal culture of its own.
It may also be a community within a society.
Present systems carry traces of this: in sentencing circles and community courts, community members take part in deciding a response, with the aim of restoring the person's place in the community rather than removing them from it.

From a \textbf{social justice} perspective, a just society is one where groups, especially marginalised ones, can voice their different needs and have those needs met -- a view that highlights inequalities between groups.
The central criticism of COMPAS came from this direction: its errors fell unevenly across racial groups.
The unevenness traced back to the system's inputs -- arrests, records, prior contacts with the police -- which carry the marks of existing disadvantage: the system took in a group's position and returned it as that group's risk.
The conception carries the criticism further: the groups bearing those errors had no voice in the system -- no part in defining the risk it measured or the needs it served.
A system designed on this basis would begin from that voice: the affected groups would take part in deciding what it measures, how it is used, and whether it is used at all.

\paragraph{Summary}
Some of the criticism of COMPAS focused on its errors: the system made erroneous predictions, and made them unevenly.
Some of the criticism reached further to argue that the system resulted in inequality -- that it was not just.
Our analysis goes further in showing how the system reflected one conception of justice, an initial, unarticulated assumption that turns into a design choice to which there are clear and well established alternatives.
This is where justice already enters design: as the first and most consequential design decision -- one that determines what is measured, who is heard, and what kind of system, if any, gets built.

\subsection{Food democracy}

The human--computer interaction community has addressed the just distribution of tangible resources such as food  and worked how to design systems for them \citep{Dombrowski2016}.
\citet{Prost2018} studied civic food networks -- community-run initiatives for buying and distributing food, coordinated through an online tool -- as a step toward food democracy, and documented how the networks struggled: sustainable food raised costs beyond what members could afford, and funding the operation raised hard questions of its own.
We take the civic food network as our case: a community distributing food among its members through a system.
Reading the case through the six conceptions suggests that these struggles were not merely practical: in them, different conceptions of justice pull the network's design in different directions.

From a \textbf{capitalist} perspective, food is the market good par excellence: produced, priced, and traded, with a just distribution resting on individual achievement and voluntary exchange.
The food democracy literature is critical of the neoliberal food system, marked by market concentration and overproduction.
The capitalist conception would read these as failures of a free market rather than consequences of one.
Accordingly, the conception steers the network toward being a better marketplace: offering organic and conventional side by side, letting prices carry the difference, and letting members' choices decide what the network provides.
When \citet{Prost2018} report a pizza-maker moving from organic to regular flour to keep costs down, this conception sees the market working: a quality on offer, a price attached, a choice made.

From a \textbf{socialist} perspective a just distribution rests on need.
What matters is not what members choose to buy but what they require:
enough, and nourishing enough, for everyone in the network.
Accordingly, the conception steers the network away from pricing altogether and toward organised provision: food distributed on need, with costs borne collectively.
Traces of this exist in the networks \citet{Prost2018} describe: some food hubs practise differentiated pricing, using wealthier members to subsidise food for poorer ones.

From a \textbf{fundamental rights} perspective, everyone has an equal right to adequate food -- a right recognised in international human rights law.
The network's task, on this conception, is to secure that right for all its members equally, whatever they can pay: no design choice, from pricing to sourcing, may cost anyone their access.
Members also hold an equal right to participate in deciding the network's rules -- a right the food democracy vision itself invokes in its call for democratic governance.
The conception thus steers the network toward guarantees: food access and an equal say secured for every member.

From a \textbf{utilitarian} perspective, a just arrangement maximises overall wellbeing, assessed impersonally.
Food translates into well-being as nutrition, and the accounting does not stop at the network's members: what the network buys, at what price, from which producers, with what environmental cost, all enters the same balance.
The conception thus steers the network by calculation: toward whatever sourcing, pricing, and provision produce the most wel-being overall.
Whether the network bakes with organic or regular flour, for example, is not a market choice but a sum: environmental gains on one side, harm to members priced out on the other.

From a \textbf{communitarian} perspective, justice is grounded in the values and practices of a particular community -- and here, unlike in the previous case, the community is already given: a civic food network is a community distributing food to itself.
What counts as good food, a fair price, or a proper way to fund the operation are for the network's own shared understanding to settle:
one network may hold local and organic sourcing as its identity, another affordability, another the relationships around food itself.
Of the six, this is the conception the food democracy vision comes closest to embodying, with its emphasis on community-run initiatives and self-governance.

From a \textbf{social justice} perspective, a just society is one where groups, especially marginalised ones, can voice their different needs and have those needs met.
Food needs differ between groups -- dietary, cultural, and religious -- and so does access to food: food poverty and insecurity fall on some groups far more than others.
On this conception, a civic food network can itself be a group's way of voicing and meeting its needs: a food-insecure community organising its own provision, on its own terms, where the surrounding food system has not met them.
The conception thus steers the design toward serving that end: a network, and a tool, that a marginalised group can run for itself.

\paragraph{Summary}
We can cast struggles \citet{Prost2018} document in a new light.
The rising costs of sustainable food, the question of how to fund the operation, the pull between affordability and quality -- each sets one conception of justice against another, and the networks were left to resolve the conflicts case by case, in practice, without the conceptions being named.
Where COMPAS reflected a single unarticulated conception, food democracy seeks to embrace several at once: a vision of community self-governance, equal rights, and needs met -- with the conflicts between them unresolved.
This is another way justice enters design: not as one assumption to be surfaced, but as several commitments already present.
Naming their conflicts turns recurring practical frustrations into questions the community can deliberate and decide.

\subsection{Gentrification}

Our third case examines how human--computer interaction researchers use justice used as an analytical lens to make sense of contemporary society and its issues.
\citet{Corbett2019} take up gentrification -- in their words, the spatial expression of economic inequality, and fundamentally a matter of social justice -- and invite HCI to engage it.
The goods most directly at stake in gentrification are space and housing: homes, rents, and who can remain in a neighbourhood.
However, the paper refocuses the question to discourses and situates justice into them:
it examines how services such as Yelp, Nextdoor, and Zillow take part in gentrification: a neighbourhood becomes ``up-and-coming'' in reviews and listings while it becomes unaffordable in fact.
We examine further how different conceptions of justice highlight aspects of the phenomena and thus guide the analytical process.

Their answers diverge more here than in our previous cases.
From a \textbf{capitalist} perspective, what is at stake is property.
Gentrification is a housing market working as markets work.
Even the discursive layer reads as a marketplace of ideas: the ticket to reviewing a restaurant, and so to describing the neighbourhood, is dining there.
From a \textbf{socialist} perspective, the stake is housing as a need: homes people depend on, lost to those who can pay more.

From a \textbf{communitarian} perspective, the stake is the neighbourhood itself -- a community's character and continuity.
From this perspective and with the framing of discourses, the question becomes which community's understanding of the place counts: the long-standing residents' or the arriving one's.
A \textbf{utilitarian} perspective keeps the whole ledger open: the welfare of the displaced and the arriving alike, counted together, impersonally.

On participation in the discourse, the case surfaces a tension between the two remaining conceptions more concretely than cases above.
From a \textbf{fundamental rights} perspective, everyone holds an equal right to take part in the discourses that construct a location.
The design question is what stands in the way: barriers of access, language, and platform design that fall on some more than others.
From a \textbf{social justice} perspective, equal access is not enough.
Where a group is under threat of displacement, its voice merits amplification: the platforms should carry the threatened residents' account of the place, not merely admit it.
The two conceptions agree that participation matters.
They disagree about power: whether a just discourse gives everyone the same voice, or more voice to those with the most at stake.

\paragraph{Summary}

\citet{Corbett2019} call to understand the lived experiences of the gentrified places the work on the map's local side, in the registers of social justice and communitarianism.
Reflecting the concern for social justice, this perspective highlights the need to understand `the lived experiences of the gentrified'.
The proposal to ``\textit{equitably} distribute placemaking'' (emphasis ours) focuses on the goods central to social justice and fundamental rights conceptions -- participation and its barriers -- instead of material outcomes concerning housing.
What equitable means is left open -- and as the map shows, the conceptions each have their own answer.


\subsection{Robots for social justice}

Our final case is neither a deployed system nor an initiative, but a framework offered to designers.
\citet{Zhu2024} present Robots for Social Justice: guidance for equitable engineering practices in human--robot interaction, building on the relation between politics and technology \citep{winner85}.
Engineering for social justice, in their definition, enhances human capabilities through the equitable distribution of opportunities and resources within specific communities.
Drawing on \citet{leydens2017engineering}, the framework is organised around understanding context and structural conditions, mobilising political power, identifying opportunities for improvement, and reducing risks and harms.
We put the map to a different use than in the previous cases:
rather than asking what system a conception would design, we read the framework itself against the map -- which conceptions of justice it reflects, and what it leaves open.
Reading the framework against the map's three questions shows what it addresses -- and what it does not.

It provides its most extensive account of \textbf{who} justice concerns.
The framework directs engineering work to specific communities, starting from their own ``constituent struggles, concerns, desires, and preferences'' \citep{Zhu2024}.
In the map's terms, the recipients are local -- the side of the map where communitarianism and social justice sit.
Elements of both are present in the framework: the communitarian appeal to a community's own preferences, and the social justice conception's vocabulary of struggles and of mobilising political power.

Less is said about what the framework identifies as \textbf{the central goods} to be distributed.
It names opportunities to be improved, for example in terms of health, education, housing, and employment, but it also lists risks and harms to be reduced.
These reflect a utilitarian focus on the elements of well-being and the reduction of harm, often shared by communitarian accounts.
However, some of these goods could also be read as building the kind of capabilities that are central to both fundamental rights and social justice conceptions.

The framework takes no position on the \textbf{basis of distribution}, even though equitable distribution stands at the heart of its definition of engineering for social justice \citep{Zhu2024}.
On the map, what is equitable depends on the basis: each conception of justice has its own understanding of an equitable distribution.
Already the map's chief distinction, between achievement and need, marks how deeply these understandings diverge -- and within each side, the conceptions differ further in their details.

\paragraph{Summary}

The lesson of this case concerns not a particular system or design, but what it takes to give a framework of justice for HCI.
The map supplies the questions such a framework needs to answer: for whom justice is sought, what goods are at stake, and on what basis they are to be distributed.
A framework need not settle these questions the same way for every context -- but if it remains silent on them, or poses inconsistent demands, it will quickly outrun its concrete usefulness.

\section{Discussion}

Human--computer interaction research has already acknowledged that justice is not a single thing \citep{Dombrowski2016}.
Drawing together our cases and conceptual work with the map, we move beyiond this observation to ask what can human–computer interaction learn, and where might it move next with justice?
First, we examine justice as something that occupies a position on the map, highlighting its pluralistic meaning and offering a way to think about justice.
Accordingly, in the next section we examine how this insight should change how we work with justice, introducing the idea of essentially contested concepts to highlight this problem.
Next, we move to the level of the research community and wider society, asking how different locations of justice should be understood.
To provide a more concrete outcome, we draw on the spiral of science and invite readers to reconsider review practices or pursue more extensive design work.

\subsection{Justice as a position}

The map -- and our cases -- suggest that justice can be seen as a position.
COMPAS was not simply lacking in justice; it stood somewhere quite specific -- near a utilitarian conception, weighing harm against harm, with recidivism prediction as the natural core of such a weighing.
The civic food networks did not fall short of food democracy's promise; instead, they stood on several conceptions at once, and their recurring struggles were the friction between them.

Our map of justice enables locating artifacts, frameworks, and problem-formulations in this manner.
This is a different activity from passing judgement on them, and it yields different things.
A judgment ends a conversation; a location begins one.
Once COMPAS is placed near utilitarianism, its famous controversy becomes legible as a disagreement between positions -- a utilitarian design met by critics standing on fundamental rights and social justice -- rather than just a technical dispute about error rates.
Once the food networks' tensions are placed, they stop being recurring practical frustrations and become questions to be addressed.
Affordability and sustainability are not just logistical problems; their conflict involves conceptions of justice pulling against each other.

The basis for this is that the map is not one of purely scholarly theories that researchers and practitioners might select among, but of \textit{live and contested conceptions} of what a just society would be. 
These conceptions are held by people, embodied in
institutions and practices, and enacted through collective action;
they may overlap, conflict, or coexist without being reducible to one another.
Our work participates in these ongoing disagreements rather than standing apart from them.

It follows that justice is not a problem a sufficiently ingenious designer could resolve:
one cannot settle, once and for all, what justice requires, because that question is not merely a technical one. 
What a designer, a framework, or a discipline can do is to become more deliberate about how it moves through the terrain: to recognize that it is positioned among a number of competing possibilities, rather than treating its own position as the neutral or self-evident one.

\subsection{How to work with essentially contested concepts?}


How can we navigate the challenge that emerges from justice being pluralistic and seeing it through a position on the map?
In the social sciences, concepts like justice (and fairness) are known as essentially contested concepts.
They carry a strong normative valence, but are at the same time complex and open in their meaning, making them open to interpretation and, as the name suggests, to contestation over their meaning \citep{Gallie1956,Collier2006}.
Classical examples of essentially contested concepts in the social sciences include 
democracy \citep{Gallie1956,Collier2006}, the Christian life \citep{Gallie1956}, social justice \citep{Gallie1956}, and the rule of law \citep{Collier2006}.
\citet{Dombrowski2016} call out the essentially contested nature of justice out in the context of human--computer interaction:
``there are many different types of social justice, no single, agreed-upon definition, and no clear consensus on how to work towards it or to verify its achievement.''
Indeed, we have above concluded here are many legitimate approaches to these concepts, inviting us to consider how human--computer interaction scholars work with justice.
This includes justice, but also other areas where human--computer interaction scholars engages with societal issues and problems, deploying and studying technology within society.


Common recommendations may include increasing reflection in the discussion and heightened self-awareness throughout the research cycle, but can we move beyond this to consider how to design with essentially contested concepts?
Design work often narrows the implementation down to a particular viewpoint, which critical research often identifies as a site of politics in technology \citep[cf.][]{winner85}.
One direction for designing with essentially contested concepts is to mimic how society already works through them.
For example, they regularly emerge in political speech and decision-making: politicians may strive towards an equal society, but equality is understood differently by different people.
In these cases, the political system — through discussions, negotiations, and ultimately voting — develops a working understanding of the concept, which is then written into laws, regulations, policies, or similar documents.
Some of these ideas are present in the jury learning approach \citep{Gordon2022}, even though it lacks the deliberation and negotiation aspects.
Alternatively, value-sensitive design \citep[e.g.,][]{Friedman2009} approaches value choices something to be examined and discussed among the people who use or are impacted by the system, often with the aim of reaching some form of shared consensus within the group.
Similar approach can be used to navigate contested concepts more directly, seeking to extract the meaning of a concept more explicitly or bring it up to guide the design work.
What seems clear is that there is \textit{no way to avoid the normative choices} embedded in these concepts.
Rather, we believe that acknowledging this contested nature is important, as it opens up space for further reflection on the implications on what is aimed and why, addressing it as an open problem.



\subsection{Justice, human--computer interaction and the society}

In response to our cases and discussion, we conclude that the notion of a \textit{just society} is a live and contested conception.
Each form of justice advances a different shape of society.
Therefore, working on justice-inspired work is also work that shapes society.
This puts more responsibility on human–computer interaction researchers working on this topic.
We first unpack the implications of this for our research community, and following that, for wider society.

While our cases illustrate the significance of different justices, they do not exhaustively account for what conceptions of justice are used in human–computer interaction.
Rather, these conceptions make visible in what gets built and studied, which harms count as findings, which communities become objects of concern, and which criticisms feel natural to make and which feel foreign.
Human--computer interaction researchers may privilege some interests or understandings over others, or make one conception of justice more consequential in practice.
However, this does not end the tensions.
We ought to ask which voices and conceptions are currently mainstream and which may be silenced in the discussion, and what implications it has for our work.

What is at stake is not only our research community: acknowledging the essentially contested nature of justice is critical due to the increasing role of computing in society \citep[e.g.,][]{Wagner2021,Burrell2021}.
Conceptions of justice are held by people, embodied in institutions and practices, and enacted through collective action; they may overlap, conflict, or coexist without being reducible to one another.
Digital interfaces and services are embedded in this social world; their design and operation participate in these ongoing disagreements rather than standing apart from them.
Therefore, every time designers and researchers settle what justice is for them, it is not only a design decision but also a political statement.
This is true for other contested concepts, such as fairness.
As we highlighted in the introduction, society is also continually settling what justice is, with its meaning shifting over time.
How do we, as a research community, engage with society in this process?

Indeed, while we take no stand concerning which conception on the map is superior, a singular focus on a particular conception of justice may lead to extremes \citep{extreme_justice}.
For example, a commitment to market fairness, taken far enough, can end up treating health, education, or political standing as things to be purchased.
A commitment to rights, taken far enough, can turn every disagreement into a legal claim, indifferent to context.
An overzealous concern with the aggregate good may trample the just treatment of individuals. A commitment to shared values, taken far enough, can turn belonging into a demand for conformity.
The details differ, but the shape of the failure is the same.
If this is right, it carries a practical implication for a field like human–computer interaction, where it may be worth asking not only which conception of justice a design decision relies on, but whether that conception has been pushed to its extremes, and how to avoid such black-and-white thinking.

\subsection{The Epistemic and Social Danger from the Spiral of Silence}


Given that the essentially contested nature of justice cannot be settled and becomes a political act, we finally reflect on the implications for the academic community.
Focusing on political scientists, \citet{Norris2021} studied whether they feel unable to fully present their perspectives when their ideological stance does not correspond with that of the dominant culture.
She found evidence of such behaviour both when the dominant culture leans politically right and when it leans politically left.
This may lead to a spiral of silence, where the minority remains less vocal about their perspectives and thus appears smaller, making it even more difficult to dissent from the dominant culture — creating a self-reinforcing loop.
The spiral of silence has been examined in citizens' discussions \citep{Soffer2017,Masullo2020,Matthes2018,Zerback2016}, but as \citet{Norris2021} shows, it also impacts academic communities.
Two challenges emerge from this for the academic community.
First, a narrow understanding of justice may limit how well we can produce knowledge about technologies and their designs: as our cases illustrate, different conceptualisations drive towards different problematisations and open (and close) different design spaces.
Second, for wider society, the challenge emerges when there is a mismatch between academic perspectives and those considered relevant by policymakers, software companies, and the general population.



Inspired by the works highlighting the dangers of the spiral of silence, we end our work by arguing for the need to create space to explore essentially contested concepts through research and design.
First, this requires open-ended examination and review practices which focus on the soundness of the argument, not its social acceptability.
Second, this invites wider use of the research tools available to us:
those engaged in design-driven justice work can present and contrast alternative designs, creating a dialogue between them and illustrate different societies.
Third, those using justice as an analytic lens need to carefully consider what the lens reveals and what it hides; as our case on gentrification highlighted, the problem may look alien from different perspectives, which may call for closer attention in the discussion and limitations sections, if not in the main analysis.
This way, we keep the academic discussion open to different conceptions and avoid the accidental creation of a dominant understanding that people do not want to disagree with.

\section{Conclusions}

Our work responds to ongoing political shifts and changing discourses about justice.
What a just society would be is under open dispute, andv human--computer interaction works inside societies having that dispute.
Drawing on political philosophy, we introduced the Map of Justice identifying six conceptions: capitalism, socialism, fundamental rights, utilitarianism, communitarianism, and social justice.
Across four cases, we showed that each already reflects a position or positions on the map, typically without the
position having been articulated, and that under other conceptions, very different systems,  analyses, problem identifications, and outcomes would follow.

We argued that justice is best treated in HCI as an explicitly plural concept -- a perspective that has received limited attention in the field's justice-oriented literature.
The point is not that any conception is wrong, but that letting one dominate sets aside considerations that other well-argued conceptions take to be central.
Justice here keeps company with essentially contested concepts: strongly normative, open in meaning, and subject to genuine and sustained disagreement.
Concepts of this kind demand a particular care in scholarly and designerly work: stating which conception is meant, and knowing what the alternatives; even seeking to examine the work across conceptualisations to embrace the dispute.
We argue that asking \textit{which justice} is a critical next step in human--computer interaction, given the changing political climate,
increased awareness on the societal impact of our work,
and the danger of extreme positions of justice.

\bibliographystyle{ACM-Reference-Format}
\bibliography{library,extra}

@article{christin2017algorithms,
  title={Algorithms in practice: Comparing web journalism and criminal justice},
  author={Christin, Ang{\`e}le},
  journal={Big data \& society},
  volume={4},
  number={2},
  pages={2053951717718855},
  year={2017},
  publisher={SAGE Publications Sage UK: London, England}
}

@inproceedings{corbett2017algorithmic,
  title={Algorithmic decision making and the cost of fairness},
  author={Corbett-Davies, Sam and Pierson, Emma and Feller, Avi and Goel, Sharad and Huq, Aziz},
  booktitle={Proceedings of the 23rd acm sigkdd international conference on knowledge discovery and data mining},
  pages={797--806},
  year={2017}
}

@book{lotter2011poverty,
  title={Poverty, ethics and justice},
  author={L{\"o}tter, Hennie},
  year={2011},
  publisher={University of Wales Press}
}

@BOOK{Walton2019-ru,
  title     = "Technical communication after the social justice turn",
  author    = "Walton, Rebecca and Moore, Kristen and Jones, Natasha",
  publisher = "Routledge",
  month     =  jun,
  year      =  2019,
  address   = "London, England",
  language  = "en"
}

@book{leydens2017engineering,
  author    = {Leydens, Jon A. and Lucena, Juan C.},
  title     = {Engineering Justice: Transforming Engineering Education and Practice},
  year      = {2017},
  publisher = {John Wiley \& Sons}
}

@article{hayry2018doctrines,
  title={Doctrines and dimensions of justice: Their historical backgrounds and ideological underpinnings},
  author={H{\"a}yry, Matti},
  journal={Cambridge Quarterly of Healthcare Ethics},
  volume={27},
  number={2},
  pages={188--216},
  year={2018},
  publisher={Cambridge University Press}
}

@article{hayry2021just,
  title={Just better utilitarianism},
  author={H{\"a}yry, Matti},
  journal={Cambridge Quarterly of Healthcare Ethics},
  volume={30},
  number={2},
  pages={343--367},
  year={2021},
  publisher={Cambridge University Press}
}

@article{rydenfelt2024climate,
  title={Climate change and transformations of justice—Views of just distribution in the Finnish policy debate},
  author={Rydenfelt, Henrik and Nyfors, Tina},
  journal={Ethics in Science and Environmental Politics},
  volume={24},
  pages={1--14},
  year={2024}
}

@article{rydenfelt2021justice,
  title={From justice to the good? Liberal utilitarianism, climate change and the coronavirus crisis},
  author={Rydenfelt, Henrik},
  journal={Cambridge Quarterly of Healthcare Ethics},
  volume={30},
  number={2},
  pages={376--383},
  year={2021},
  publisher={Cambridge University Press}
}

@article{hayry2025oikeudenmukainen,
  title={Oikeudenmukainen siirtym{\"a} eettiseen kest{\"a}vyyteen: ei miksi ja mit{\"a} vaan miten},
  author={H{\"a}yry, Matti and Takala, Tuija-Maija},
  journal={Ajatus},
  number={81},
  pages={13--18},
  year={2025},
  publisher={Suomen Filosofinen Yhdistys}
}

@book{locke,
  title={An essay concerning human understanding},
  author={Locke, John},
  year={1689}
}

@book{nozick1974anarchy,
  title={Anarchy, State, and Utopia},
  author={Nozick, Robert},
  publisher={Basic Books},
  year={1974}
}

@book{marx,
  title= {Das Kapital. Kritik der politischen Ökonomie},
  author= {Marx, Karl},
  year= {1867},
  publisher= {Verlag von Otto Meissner}
}

@article{gilabert2015socialist,
  title={The socialist principle “from each according to their abilities, to each according to their needs”},
  author={Gilabert, Pablo},
  year={2015}
}

@book{rawls1971theory,
  author = {Rawls, John},
  title = {A theory of justice},
  publisher = {Harvard University Press},
  year = {1971}
}

@book{bentham,
  author = {Jeremy Bentham},
  title = {A Fragment on Government. In An Introduction to the Principles of Morals and Legislation},
  year = {1789}
}

@book{mill,
  author = {John Stuart Mill},
  title = {Utilitarianism},
  year = {1861}
}

@book{sandel,
  author = {Sandel, Michael J.},
  title = {Liberalism and the Limits of Justice,},
  publisher = {Cambridge University Press},
  year = {1981}
}

@book{walzer,
  author = {Walzer, Michael},
  publisher = {Blackwell},
  title = {Spheres of Justice: A Defense of Pluralism and Equality},
  year = {1983}
}

@book{gilligan1982different,
  author    = {Gilligan, Carol},
  title     = {In a Different Voice: Psychological Theory and Women's Development},
  year      = {1982},
  publisher = {Harvard University Press}
}

@book{young1990justice,
  author    = {Young, Iris Marion},
  title     = {Justice and the Politics of Difference},
  year      = {1990},
  publisher = {Princeton University Press},
}

@article{fraser2000rethinking,
  author  = {Fraser, Nancy},
  title   = {Rethinking Recognition},
  journal = {New Left Review},
  year    = {2000},
  volume  = {3},
  pages   = {107--120}
}

@techreport{hyvinvointivalitio,
    author = {Rydefelt, Henrik and Jensen-Eriksen, Niklas and Lundstendt, Tero},
    title = {Hyvinvointivaltion alavire. Nykyisen yhteiskuntajärjestyksen historia},
    institution = {Libera},
    year = {2024} 
}

@book{aristotle,
  title={Nicomachean ethics},
  author={Aristotle},
  year={350 B.C.E },
  publisher={https://classics.mit.edu/Aristotle/nicomachaen.html}
}

@misc{angwin2016machine,
  author       = {Angwin, Julia and Larson, Jeff and Mattu, Surya and Kirchner, Lauren},
  title        = {Machine Bias},
  howpublished = {ProPublica},
  year         = {2016},
  url          = {https://www.propublica.org/article/machine-bias-risk-assessments-in-criminal-sentencing},
}

@inbook{extreme_justice,
    title = "Extreme Justice",
    author = "Henrik Rydenfelt",
    year = "2026",
    series = "Acta Philosophica Fennica",
    publisher = "Philosophical Society of Finland",
    pages = "133--147",
    editor = "Henrik Rydenfelt and Tuija Takala",
    booktitle = "Justice to Extinction",
}

@book{Nussbaum,
 author = {Martha C. Nussbaum},
 publisher = {Harvard University Press},
 title = {Frontiers of Justice: Disability, Nationality, Species Membership},
 urldate = {2026-09-10},
 year = {2006}
}

@book{sen2009idea,
  author    = {Sen, Amartya},
  title     = {The Idea of Justice},
  publisher = {Harvard University Press},
  address   = {Cambridge, MA},
  year      = {2009}
}

@incollection{ubuntu,
	author = {M. B. Ramose},
	booktitle = {Philosophy from Africa: A text with readings},
	editor = {P. H. Coetzee and A. P. J. Roux},
	publisher = {Oxford University Press South Africa},
	title = {The Philosophy of Ubuntu as a Philosophy},
	year = {2003}
}

@book{Collins2000,
  author    = {Collins, Patricia Hill},
  title     = {Black Feminist Thought: Knowledge, Consciousness, and the Politics of Empowerment},
  edition   = {2nd},
  publisher = {Routledge},
  address   = {New York},
  year      = {2000}
}

@article{Crenshaw1991,
  author  = {Crenshaw, Kimberl{\'e} Williams},
  title   = {Mapping the Margins: Intersectionality, Identity Politics, and Violence against Women of Color},
  journal = {Stanford Law Review},
  volume  = {43},
  number  = {6},
  pages   = {1241--1299},
  year    = {1991},
  doi     = {10.2307/1229039}
}

@book{Young1990,
  author    = {Young, Iris Marion},
  title     = {Justice and the Politics of Difference},
  publisher = {Princeton University Press},
  address   = {Princeton, NJ},
  year      = {1990}
}

@inproceedings{Dombrowski2016,
address = {New York, NY, USA},
author = {Dombrowski, Lynn and Harmon, Ellie and Fox, Sarah},
booktitle = {Proceedings of the 2016 ACM Conference on Designing Interactive Systems},
doi = {10.1145/2901790.2901861},
isbn = {9781450340311},
month = {jun},
pages = {656--671},
publisher = {ACM},
title = {{Social Justice-Oriented Interaction Design}},
url = {https://dl.acm.org/doi/10.1145/2901790.2901861},
year = {2016}
}

@article{Mulligan2019,
author = {Mulligan, Deirdre K and Kroll, Joshua A and Kohli, Nitin and Wong, Richmond Y},
doi = {10.1145/3359221},
issn = {2573-0142},
journal = {Proceedings of the ACM on Human-Computer Interaction},
month = {nov},
number = {CSCW},
pages = {1--36},
publisher = {CSCW},
title = {{This Thing Called Fairness}},
url = {https://doi.org/10.1145/3359221 https://dl.acm.org/doi/10.1145/3359221},
volume = {3},
year = {2019}
}

@article{Norris2021,
author = {Norris, Pippa},
doi = {10.1177/00323217211037023},
issn = {0032-3217},
journal = {Political Studies},
month = {feb},
number = {1},
pages = {145--174},
title = {{Cancel Culture: Myth or Reality?}},
url = {http://journals.sagepub.com/doi/10.1177/00323217211037023},
volume = {71},
year = {2023}
}

@article{Collier2006,
author = {Collier, David and Hidalgo, Fernando Daniel and Maciuceanu, Andra Olivia},
doi = {10.1080/13569310600923782},
issn = {13569317},
journal = {Journal of Political Ideologies},
number = {3},
pages = {211--246},
title = {{Essentially contested concepts: Debates and applications}},
volume = {11},
year = {2006}
}

@article{Masullo2020,
author = {Masullo, Gina M. and Lu, Shuning and Fadnis, Deepa},
doi = {10.1177/1461444820954194},
issn = {14617315},
journal = {New Media and Society},
title = {{Does online incivility cancel out the spiral of silence? A moderated mediation model of willingness to speak out}},
year = {2020}
}

@article{Ohlund2025,
author = {{\"{O}}hlund, Linnea and Wiberg, Mikael},
doi = {10.1093/iwc/iwaf009},
issn = {09535438},
journal = {Interacting with Computers},
number = {6},
pages = {553--567},
title = {{Social Justice in HCI: Current Streams, Considerations, and Ways Forward}},
volume = {37},
year = {2025}
}

@article{Bellini2022,
author = {Bellini, Rosanna and Leal, Debora De Castro and Dixon, Hazel Anneke and Fox, Sarah E. and Strohmayer, Angelika},
doi = {10.1145/3491101.3503698},
isbn = {9781450391566},
journal = {Conference on Human Factors in Computing Systems - Proceedings},
title = {{"There is no justice, just us": Making mosaics of justice in social justice Human-Computer Interaction}},
year = {2022}
}

@article{Gallie1956,
author = {Gallie, W. B.},
doi = {10.1093/aristotelian/56.1.167},
issn = {0066-7374},
journal = {Proceedings of the Aristotelian Society},
month = {jun},
number = {1},
pages = {167--198},
title = {{IX.—Essentially Contested Concepts}},
url = {https://academic.oup.com/aristotelian/article-lookup/doi/10.1093/aristotelian/56.1.167},
volume = {56},
year = {1956}
}

@inproceedings{Paakkonen2020,
address = {New York, NY, USA},
author = {P{\"{a}}{\"{a}}kk{\"{o}}nen, Juho and Nelimarkka, Matti and Haapoja, Jesse and Lampinen, Airi},
booktitle = {Proceedings of the 2020 CHI Conference on Human Factors in Computing Systems},
doi = {10.1145/3313831.3376780},
isbn = {9781450367080},
month = {apr},
pages = {1--14},
publisher = {ACM},
title = {{Bureaucracy as a Lens for Analyzing and Designing Algorithmic Systems}},
url = {https://dl.acm.org/doi/10.1145/3313831.3376780},
year = {2020}
}

@article{Matthes2018,
author = {Matthes, J{\"{o}}rg and Knoll, Johannes and von Sikorski, Christian},
doi = {10.1177/0093650217745429},
isbn = {0093-6502},
issn = {15523810},
journal = {Communication Research},
number = {1},
pages = {3--33},
title = {{The “Spiral of Silence” Revisited: A Meta-Analysis on the Relationship Between Perceptions of Opinion Support and Political Opinion Expression}},
volume = {45},
year = {2018}
}

@article{Lipton2017,
archivePrefix = {arXiv},
arxivId = {1711.07076},
author = {Lipton, Zachary C. and Chouldechova, Alexandra and McAuley, Julian},
eprint = {1711.07076},
pages = {1--19},
title = {{Does mitigating ML's impact disparity require treatment disparity?}},
url = {http://arxiv.org/abs/1711.07076},
year = {2017}
}

@inproceedings{Alkhatib2019,
address = {New York, New York, USA},
author = {Alkhatib, Ali and Bernstein, Michael},
booktitle = {Proceedings of the 2019 CHI Conference on Human Factors in Computing Systems - CHI '19},
doi = {10.1145/3290605.3300760},
isbn = {9781450359702},
pages = {1--13},
publisher = {ACM Press},
title = {{Street-Level Algorithms}},
url = {http://dl.acm.org/citation.cfm?doid=3290605.3300760},
year = {2019}
}

@inproceedings{Boone2024,
address = {New York, NY, USA},
author = {Chordia, Ishita and Baltaxe-Admony, Leya Breanna and Boone, Ashley and Sheehan, Alyssa and Dombrowski, Lynn and {Le Dantec}, Christopher A and Ringland, Kathryn E and Smith, Angela D R},
booktitle = {Proceedings of the CHI Conference on Human Factors in Computing Systems},
doi = {10.1145/3613904.3642704},
isbn = {9798400703300},
month = {may},
number = {1},
pages = {1--33},
publisher = {ACM},
title = {{Social Justice in HCI: A Systematic Literature Review}},
url = {https://dl.acm.org/doi/10.1145/3613904.3642704},
volume = {1},
year = {2024}
}

@inproceedings{Kasy2021,
address = {New York, NY, USA},
author = {Kasy, Maximilian and Abebe, Rediet},
booktitle = {Proceedings of the 2021 ACM Conference on Fairness, Accountability, and Transparency},
doi = {10.1145/3442188.3445919},
isbn = {9781450383097},
month = {mar},
pages = {576--586},
publisher = {ACM},
title = {{Fairness, Equality, and Power in Algorithmic Decision-Making}},
url = {https://dl.acm.org/doi/10.1145/3442188.3445919},
volume = {11},
year = {2021}
}

@incollection{Friedman2009,
address = {Hoboken, NJ, USA},
author = {Friedman, Batya and Kahn, Peter H. and Borning, Alan},
booktitle = {The Handbook of Information and Computer Ethics},
doi = {10.1002/9780470281819.ch4},
isbn = {9780471799597},
pages = {69--101},
publisher = {John Wiley {\&} Sons, Inc.},
title = {{Value Sensitive Design and Information Systems}},
url = {http://doi.wiley.com/10.1002/9780470281819.ch4},
year = {2009}
}

@article{Burrell2021,
author = {Burrell, Jenna and Fourcade, Marion},
doi = {10.1146/annurev-soc-090820-020800},
issn = {0360-0572},
journal = {Annual Review of Sociology},
month = {jul},
number = {1},
pages = {annurev--soc--090820--020800},
title = {{The Society of Algorithms}},
url = {https://www.annualreviews.org/doi/10.1146/annurev-soc-090820-020800},
volume = {47},
year = {2021}
}

@article{Burrell2024,
author = {Burrell, Jenna},
doi = {10.5210/fm.v29i4.13630},
issn = {1396-0466},
journal = {First Monday},
month = {apr},
title = {{Automated decision-making as domination}},
url = {https://firstmonday.org/ojs/index.php/fm/article/view/13630},
year = {2024}
}

@article{Corbett-Davies2018,
archivePrefix = {arXiv},
arxivId = {1808.00023},
author = {Corbett-Davies, Sam and Goel, Sharad},
eprint = {1808.00023},
journal = {arXiv},
month = {jul},
number = {Ec},
title = {{The Measure and Mismeasure of Fairness: A Critical Review of Fair Machine Learning}},
url = {http://arxiv.org/abs/1808.00023},
year = {2018}
}

@inproceedings{Zhu2024,
address = {New York, NY, USA},
author = {Zhu, Yifei and Wen, Ruchen and Williams, Tom},
booktitle = {Proceedings of the 2024 ACM/IEEE International Conference on Human-Robot Interaction},
doi = {10.1145/3610977.3634944},
isbn = {9798400703225},
issn = {21672148},
month = {mar},
pages = {850--859},
publisher = {ACM},
title = {{Robots for Social Justice (R4SJ): Toward a More Equitable Practice of Human-Robot Interaction}},
url = {https://dl.acm.org/doi/10.1145/3610977.3634944},
year = {2024}
}

@article{Corbett2019,
author = {Corbett, Eric and Loukissas, Yanni},
doi = {10.1145/3290605.3300510},
isbn = {9781450359702},
journal = {Conference on Human Factors in Computing Systems - Proceedings},
pages = {1--16},
title = {{Engaging gentrification as a social justice issue in HCI}},
year = {2019}
}

@article{Prost2018,
author = {Prost, Sebastian and Crivellaro, Clara and Haddon, Andy and Comber, Rob},
doi = {10.1145/3173574.3173907},
isbn = {9781450356206},
journal = {Conference on Human Factors in Computing Systems - Proceedings},
pages = {1--14},
title = {{Food democracy in the making: Designing with local food networks}},
volume = {2018-April},
year = {2018}
}

@inproceedings{Selbst2019,
address = {New York, NY, USA},
author = {Selbst, Andrew D. and Boyd, Danah and Friedler, Sorelle A. and Venkatasubramanian, Suresh and Vertesi, Janet},
booktitle = {Proceedings of the Conference on Fairness, Accountability, and Transparency},
doi = {10.1145/3287560.3287598},
isbn = {9781450361255},
month = {jan},
pages = {59--68},
publisher = {ACM},
title = {{Fairness and Abstraction in Sociotechnical Systems}},
url = {https://dl.acm.org/doi/10.1145/3287560.3287598},
year = {2019}
}

@inproceedings{Gordon2022,
address = {New York, NY, USA},
archivePrefix = {arXiv},
arxivId = {2202.02950},
author = {Gordon, Mitchell L. and Lam, Michelle S. and Park, Joon Sung and Patel, Kayur and Hancock, Jeff and Hashimoto, Tatsunori and Bernstein, Michael S.},
booktitle = {CHI Conference on Human Factors in Computing Systems},
doi = {10.1145/3491102.3502004},
eprint = {2202.02950},
isbn = {9781450391573},
month = {apr},
pages = {1--19},
publisher = {ACM},
title = {{Jury Learning: Integrating Dissenting Voices into Machine Learning Models}},
url = {https://dl.acm.org/doi/10.1145/3491102.3502004},
year = {2022}
}

@inproceedings{Marshall2017,
address = {New York, New York, USA},
author = {Marshall, Joe and Linehan, Conor and Spence, Jocelyn C. and {Rennick Egglestone}, Stefan},
booktitle = {Proceedings of the 2017 CHI Conference Extended Abstracts on Human Factors in Computing Systems - CHI EA '17},
doi = {10.1145/3027063.3052752},
isbn = {9781450346566},
pages = {848--857},
publisher = {ACM Press},
title = {{A Little Respect: Four Case Studies of HCI's Disregard for Other Disciplines}},
url = {http://dl.acm.org/citation.cfm?doid=3027063.3052752},
year = {2017}
}

@article{Wagner2021,
author = {Wagner, Claudia and Strohmaier, Markus and Olteanu, Alexandra and Kıcıman, Emre and Contractor, Noshir and Eliassi-Rad, Tina},
doi = {10.1038/s41586-021-03666-1},
issn = {0028-0836},
journal = {Nature},
month = {jul},
number = {7866},
pages = {197--204},
title = {{Measuring algorithmically infused societies}},
url = {http://www.nature.com/articles/s41586-021-03666-1},
volume = {595},
year = {2021}
}

@incollection{winner85,
address = {Buckingham},
author = {Winner, Langdon},
booktitle = {The social shaping of technology},
editor = {MacKenzie, Donald and Wajcman, Judy},
pages = {26--38},
publisher = {Open University Press},
title = {{Do artifacts have politics?}},
year = {1985}
}

@article{Soffer2017,
author = {Soffer, Oren and Gordoni, Galit},
doi = {10.1080/1369118X.2017.1281991},
issn = {14684462},
journal = {Information Communication and Society},
number = {October},
pages = {1--16},
title = {{Opinion expression via user comments on news websites: analysis through the perspective of the spiral of silence}},
volume = {4462},
year = {2017}
}

@article{Zerback2016,
author = {Zerback, T. and Fawzi, N.},
doi = {10.1177/1461444815625942},
issn = {1461-4448},
journal = {New Media {\&} Society},
title = {{Can online exemplars trigger a spiral of silence? Examining the effects of exemplar opinions on perceptions of public opinion and speaking out}},
url = {http://nms.sagepub.com/cgi/doi/10.1177/1461444815625942},
year = {2016}
}

\end{document}